\documentclass[journal]{IEEEtranTIE}
\usepackage{graphicx}
\usepackage{cite}
\usepackage{picinpar}
\usepackage{amsmath}
\usepackage{url}
\usepackage{flushend}
\usepackage[latin1]{inputenc}
\usepackage{colortbl}
\usepackage{soul}
\usepackage{multirow}
\usepackage{pifont}
\usepackage{color}
\usepackage{alltt}
\usepackage[hidelinks]{hyperref}
\usepackage{enumerate}
\usepackage{breakurl}
\usepackage{epstopdf}
\usepackage{pbox}

\usepackage{orcidlink}

\usepackage[nolist,nohyperlinks]{acronym}

\begin{acronym}[ECU]

\acro{AWMPS}[AWMPS]{arbitrary waveform magnetic particle spectrometer}
\acro{ADC}[ADC]{analog-to-digital converter} 
\acro{AUC}[AUC]{area under the curve}

\acro{CT}[CT]{computed tomography}

\acro{DAC}[DAC]{digital-to-analog converter}
\acro{DFG}[DFG]{drive-field generator}
\acro{DFCs}[DFCs]{drive-field coils}
\acro{DF}[DF]{drive-field}
\acro{DA}[DA]{differential amplifier}
\acro{DLS}[DLS]{dynamic light scattering}
\acro{DTS}[DTS]{dispersion technology software}
\acro{DPDT}[DPDT]{double pole double throw}

\acro{ECD}[ECD]{equivalent circuit diagram}
\acro{EMR}[EMR]{electromagnetic relay}

\acro{FFL}[FFL]{field-free-line}
\acro{FFP}[FFP]{field-free-point}
\acro{FFR}[FFR]{field-free-region}
\acro{FFT}[FFT]{fast Fourier transform}
\acro{FOV}[FOV]{field-of-view}
\acro{FWHM}[FWHM]{full width at half maximum}
\acro{FPGA}[FPGA]{field programmable gate array}

\acro{GUI}[GUI]{graphical user interface}

\acro{HCC}[HCC]{high current circuit}

\acro{ICN}[ICN]{inductive coupling network}
\acro{ISI}[ISI]{integrated signal intensity}
\acro{ICs}[ICs]{integrated circuits}
\acro{IA}[IA]{instrumentation amplifier}
\acro{ICU}[ICU]{intensive care unit} 

\acro{LC}[LC]{inductor-capacitor}
\acro{LFR}[LFR]{low-field-region}
\acro{LNA}[LNA]{low noise amplifier}

\acro{MPI}[MPI]{magnetic particle imaging}
\acro{MRI}[MRI]{magnetic resonance imaging}
\acro{MTT}[MTT]{mean-transit-time}
\acro{MNP}[MNP]{magnetic nanoparticle}
\acro{MPS}[MPS]{magnetic particle spectroscopy}
\acro{MAE}[MAE]{mean absolute error}
\acro{MSE}[MSE]{mean squared error}
\acro{MMR}[MMR]{magneto-mechanical resonator}

\acro{PA}[PA]{power amplifier}
\acro{PSF}[PSF]{point spread function}
\acro{PNS}[PNS]{peripheral nerve stimulation}
\acro{PTT}[PTT]{pulmonary transit time}
\acro{PDI}[PDI]{polydispersity index}
\acro{PCB}[PCB]{printed circuit board}
\acro{PWM}[PWM]{pulse-width modulation}

\acro{Q}[Q-factor]{quality factor}

\acro{RF}[RF]{radio-frequency}
\acro{RP}[RP]{RedPitaya STEMlab 125-14}
\acro{RF-fields}[RF]{radio frequency fields}
\acro{rBV}[rBV]{relative blood-volume}
\acro{rBF}[rBF]{relative blood-flow}
\acro{rCBV}[rCBV]{relative cerebral-blood-volume}
\acro{rCBF}[rCBF]{relative cerebral-blood-flow}
\acro{RBCs}[RBCs]{red blood cells}

\acro{SNR}[SNR]{signal-to-noise ratio}
\acro{SU}[SU]{surveillance unit}
\acro{SAR}[SAR]{specific absorption rate}
\acro{SPIONs}[SPIONs]{superparamagnetic iron-oxide nanoparticles}
\acro{USPIONs}[USPIONs]{ultrasmall superparamagnetic iron-oxide nanoparticles}
\acro{SF}{selection field}
\acro{SFG}[SFG]{selection-field generator}
\acro{SEM}[SEM]{scanning electron microscope}
\acro{SSIM}[SSIM]{structural similarity index measure}
\acro{SMD}[SMD]{surface-mount device}

\acro{TF}[TF]{transfer function}
\acro{THD}[THD]{total harmonic distortion}
\acro{TTP}[TTP]{time-to-peak}
\acro{TEM}[TEM]{transmission electron microscopy}
\acro{TxRx}[TxRx]{transmit-receive}

\acro{USPIO}[USPIO]{ultra-small superparamagnetic iron oxide}

\acro{VOI}[VOI]{volume of interest}
\acro{VSM}[VSM]{vibrating sample magnetometry}

\end{acronym}

\usepackage[exponent-product = \cdot]{siunitx}
\usepackage[font={footnotesize}]{caption}

\begin{document}
\linespread{0.93}

\title{\huge Low-cost analog signal chain for transmit-receive circuits of passive induction-based resonators}

\author{
	
	Fabian Mohn~\orcidlink{0000-0002-9151-9929}, 
	Florian Thieben~\orcidlink{0000-0002-2890-5288}, and 
    Tobias Knopp~\orcidlink{0000-0002-1589-8517}
    \vspace{-0.3cm}

	\thanks{
		All authors are with the Institute for Biomedical Imaging, Hamburg University of Technology, Hamburg, Germany and with the Section for Biomedical Imaging, University Medical Center Hamburg-Eppendorf, Hamburg, Germany (e-mail: fabian.mohn@tuhh.de). 
		
		T.K. is also with the Fraunhofer Research Institution for Individualized and Cell-Based Medical Engineering IMTE, L\"ubeck, Germany.
	}
}

\maketitle

\begin{abstract}
Passive wireless sensors are crucial in modern medical and industrial settings to monitor procedures and conditions. We demonstrate a circuit to inductively excite passive resonators and to conduct their decaying signal response to a low noise amplifier. Two design variations of a generic transmit-receive signal chain are proposed, measured, and described in detail for the purpose of facilitating replication. Instrumentation and design aim to be scalable for multi-channel array configurations, using either off-the-shelf class-D audio amplifiers or a custom full H-bridge. Measurements are conducted on miniature magneto-mechanical resonators in the ultra low frequency range to enable sensing and tracking applications of such devices in different environments.
\end{abstract}


\markboth{}%
{}

\definecolor{limegreen}{rgb}{0.2, 0.8, 0.2}
\definecolor{forestgreen}{rgb}{0.13, 0.55, 0.13}
\definecolor{greenhtml}{rgb}{0.0, 0.5, 0.0}

\section{Introduction}
\label{sec:intro}

In engineering, healthcare, and science, sensors play an indispensable role in monitoring, controlling, securing, and optimizing systems and procedures. The demand for advanced sensing technologies in harsh environments and process conditions motivates the development of low cost, wireless and preferably passive sensors of small size. \Acp{MMR} and inductor-capacitor (LC) resonators are evolving to fill this gap by responding to physical inputs from the environment such as temperature, pressure or chemical composition, translating these signals into readable output data for sensing applications and being remotely trackable~\cite{gleich_miniature_2023,fischer_miniature_2024} or suitable for actuation~\cite{fischer_magneto-oscillatory_2024}. 
The development of medical LC resonators for sensing dates back to the 1960s~\cite{collins_miniature_1967} (passive), and magneto-mechanical devices are currently used in ultra low frequency radio communications~\cite{thanalakshme_magneto-mechanical_2022}.

The concept of remote tracking of miniature \acp{MMR} has only recently been demonstrated and these devices can be cheaply manufactured, miniaturized and used in parallel~\cite{gleich_miniature_2023}. Tailored to a physical parameter that causes a change in distance or magnetization, the parameter can be coupled to the natural resonance frequency of the torsional oscillator, which can be extracted from the measured signals. Data acquisition relies on ultra low frequency magnetic fields for actuation and signal detection without the need for optical access. \acp{MMR} can be sensed and tracked using a simple signal acquisition system that alternates between transmit (Tx) and receive windows (Rx), to continuously pump the torsional oscillation and process the read out and re-excitation in real-time~\cite{gleich_miniature_2023}. Analogously, this has also been demonstrated for an LCQ sensor, which consists of an LC circuit coupled to a quartz oscillator to improve the quality factor~\cite{gleich_miniature_2023}.

The concrete implementation of an \ac{MMR} suitable \ac{TxRx} circuit 
has not been described in~\cite{gleich_miniature_2023} and the purpose of the present note is to fill this gap and propose different circuit solutions that can be utilized for inductive sensing and tracking applications. By using low-cost components, we focus on scalability to multi-coil array configurations. One type of \ac{TxRx} switch is based on an \ac{EMR} and an off-the-shelf class-D audio amplifier (\textit{Type 1}), that is mainly indented for frequencies below and in the low kHz region~\cite{knopp_empirical_2024}.
In order to achieve higher resonator frequencies and greater amplitudes, a custom full-H-bridge circuit is proposed (\textit{Type 2}), also reducing the overall instrumentation complexity. The merits and disadvantages of these approaches are evaluated, measured and compared.

\begin{figure*}[ht!]
    \centering
    \includegraphics[width=1.0\linewidth]{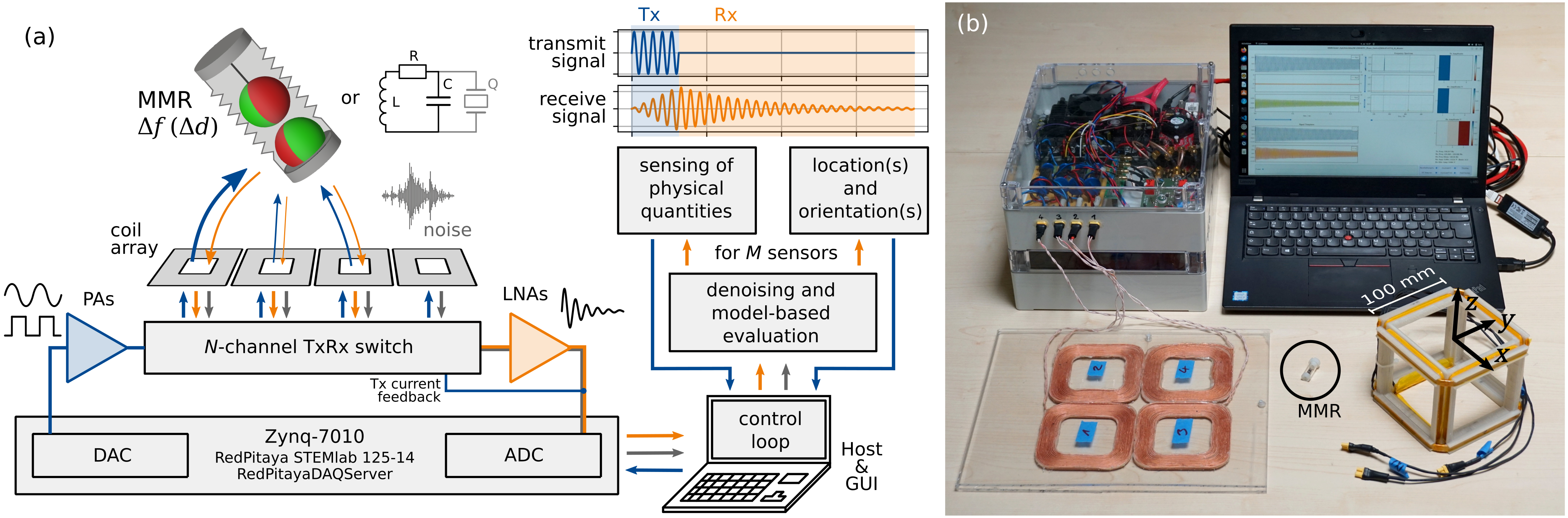}
    \caption{\textbf{Overview of MMR signal chain.} The principal components required to drive, sense and locate MMR or LC resonators are shown in (a). The depicted sensor responds to pressure changes in the environment by decreasing the magnet-to-magnet distance, thereby increasing the measured resonant frequency. Evaluation is performed in real-time on a computer, that also controls the amplitude, phase and frequency of the next transmit cycle to continuously pump the oscillator. In (b), an image of the prototype device including class-D amplifier, TxRx-switch and data acquisition system is shown. Different coils, such as a planar $4$-element coil array or the 3D square-shape Helmholtz coil can be connected and exchanged. \vspace{-0.3cm}}
    \label{fig:overview}
\end{figure*}

\section{Problem Statement}
\label{sec:problem}
A generic signal chain is illustrated in \autoref{fig:overview}\,(a), composed of an $N$ channel coil array, and a dedicated \ac{PA}, the \ac{TxRx} switch, and a \ac{LNA} for each channel. The objective is to excite the \ac{MMR} with a power signal and then receive their weak signal decay. Hence, simultaneous transmission and reception is not required. However, a parallel connection of the Tx and Rx paths could damage the sensitive \ac{LNA} electronics, which necessitates the use of a \ac{TxRx} switch for proper isolation. Due to \acp{MMR} with individual resonance frequencies and sensing applications, broadband signal amplification is required.

The challenge for the \ac{TxRx} switch lies in the fact that contemporary class-D amplifiers may utilize a full-bridge output stage, which is capable of delivering four times the output power for a given supply voltage in comparison to a half-bridge~\cite{kwek_analysis_2008,tang_design_2020}. 
Such designs are intended for use with electrically isolated loudspeakers, not an inductive single-ended coil array or single-ended \ac{LNA} design. Ground loops created between the input and output of the amplifier channels introduce a second impedance in parallel with the amplifier's internal feedback loop. As a result, the entire circuit is prone to instability, oscillations, and high bias currents.

Nevertheless, the advantages in size, output power, power efficiency and cost of class-D amplifiers are well understood~\cite{tang_design_2020,niklaus_100_2023, chang_analysis_2000, jockram_concept_2012} and exceed those of the majority of other amplification options, thereby justifying the design of a \ac{TxRx} switch capable of switching both output lines simultaneously to incorporate an off-the-shelf class-D amplifier. The issue of protecting sensitive circuit elements from the effects of hot switching needs to be addressed, given the high voltage generated by interrupting a current in a coil with high inductance. 

To cope with the typical cut-off frequency around 20\,kHz of class-D audio amplifiers and to maximize power, a simple full-bridge circuit without \ac{PWM} or low-pass filtering can combine amplification and a \ac{TxRx} switch for higher frequencies and sequence cycle times. Such a bridge is limited to pulsing currents and results in a loss of control over the excitation waveform shape and amplitude. To regain control of the Tx signal and thus of the MMR amplitude during a fixed Tx window length, a duty cycle can be implemented that can be changed during measurements.

\section{Methods}
\label{sec:methods}

Two circuit types are developed, used for MMR measurements and compared. 
An overview of the signal chain including the main components for excitation, readout, control, evaluation and parameters sensing is given in \autoref{fig:overview}.

\subsection{Type 1: Class-D and EMR switch}
\label{sec:methods:type1}
\begin{figure}[b!]
    \centering
    \includegraphics[width=1.0\linewidth]{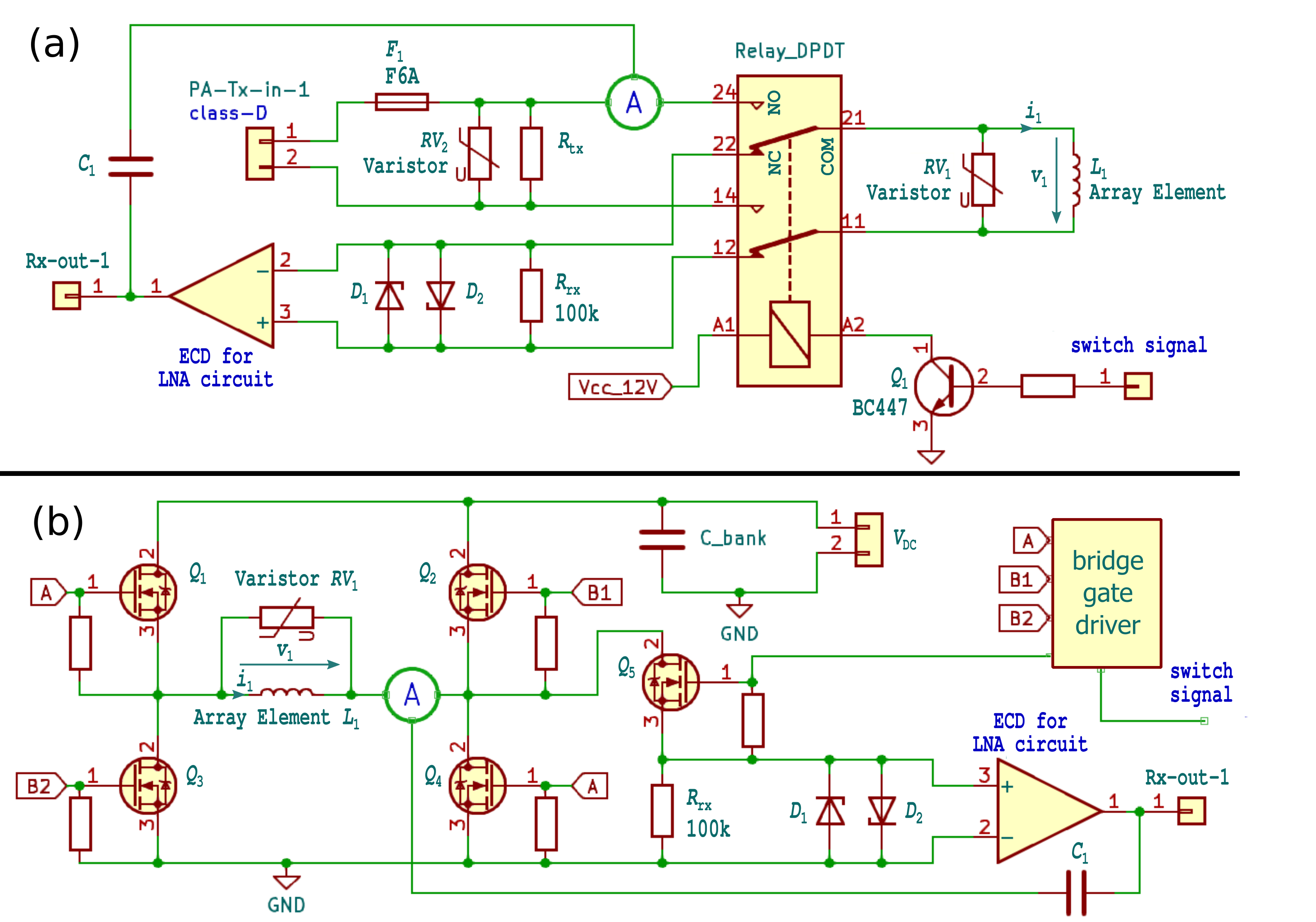}
    \caption{\textbf{Two types of TxRx switches.} An EMR relay with a \ac{DPDT} configuration is used in \textit{type 1 }(a). The COM contacts connect to the coil array $L_1$ and a class-D amplifier with a full-bridge output stage. In (b), a different design is proposed named \textit{type 2}, based on an H-bridge without PWM or filtering, combining amplifier and TxRx switch into one unit.}
    \label{fig:ECD}
\end{figure}

A \ac{DPDT} \ac{EMR} is used to switch both output lines of a channel simultaneously, acting as the \ac{TxRx} switch. This \ac{TxRx} switch can be used with any amplifier type. The array coil element $L_1$ connects to the COM (common) contact pins to ensure a consistent connection to either the amplifier during the Tx window or to the \ac{LNA} during the Rx window. 
An \acl{ECD} is shown in \autoref{fig:ECD}\,(a). The employed monostable \ac{DPDT} relay operates at \SI{12}{\volt} for switching currents of up to \SI{8}{\A} at \SI{250}{\V} (max. operation time of \SI{15}{\ms}, max. release time of \SI{8}{\ms}).

The varistors, especially $RV_1$, are dimensioned to have a clamping voltage around twice the expected coil voltage $v_1$ during transmission. Their purpose is to remain at a high impedance during normal operation, and to become conductive when the driving coil current $i_1$ is suddenly interrupted by (hot) switching, causing a high voltage over the self-inductance of $L_1$ as given by $v_1(t) = L_1 \tfrac{\partial i_1}{\partial t}$. $RV_1$ protects the \ac{EMR} and the \ac{LNA} input when switching from Tx to Rx. Diodes $D_1$ and $D_2$ are further \ac{LNA} input protection. $RV_2$ and $R_\text{tx}$ can be dimensioned as a high-impedance load or according to stability criteria of the \ac{PA}.

The implementation of an AC current sensor enables feedback of the drive current $i_1$ during transmission - using the idle Rx input. To this end, a sensor of type HO 8-NP (LEM Holding AG, Switzerland) is incorporated on the \ac{PCB} layout which is silenced during reception.

\subsection{Type 2: H-bridge amplifier and switch}
\label{sec:methods:type2}
In order to increase switching time, excitation frequency and amplitude, an H-bridge circuit is utilized in place of both the class-D amplifier and the \ac{EMR} \ac{TxRx} switch. To this end, a fifth MOSFET is incorporated in addition to the four N-channel MOSFETs of the bridge. This approach is illustrated in \autoref{fig:ECD}\,(b). It is recommended that isolated gate drivers be employed for this purpose. In this version we use MOSFETs of type IPP034N08N5 (Infineon Technologies AG, Germany). The \ac{TxRx} switch is realized by $Q_3$ and $Q_5$, which are conducting during the Rx window, while $Q_1$, $Q_2$, and $Q_4$ are non-conducting. The induced Rx voltage in $L_1$ from the resonator drops across $R_\text{rx}$ and can then be amplified. An isolated gate driver delivers the required gate-source voltages, prohibits shoot-through conditions and enables $Q_3$ during Rx. 

The limiting factor to steep slew rates is the dominant coil impedance $Z_1=R_1 +j\omega L_1$ with time constant $\tau=L_1 R_1^{-1}$ and the bridge voltage $V_\text{DC}$. The pulsed current asymptotically approaches its value with $i_1(t)=V_\text{DC} R_1^{-1} (1-e^{-t/\tau})$. Typically, the bridge voltage remains constant and the resonator amplitude can be controlled indirectly using either the number of Tx periods or the Tx duty cycle to achieve the desired variation in excitation.

\subsection{Measurement Setup} 
\label{sec:methods:setup}

A tri-axial Helmholtz-like square-shape coil, shown in \autoref{fig:overview}\,(b) on the bottom right, is used for field generation and reception due to its near-homogeneous field distribution~\cite{hurtado-velasco_simulation_2016}, simple construction and good accessibility of the enclosed volume. Measurements are taken with the $x$-channel ($L_1=\SI{730}{\micro\henry}$, $R_1=\SI{3.38}{\ohm}$). The coil side lengths are equal with \SI{95}{\mm}, the distance of the coil pair centers is \SI{107}{\mm}, and the available free space inside is approximately \SI{9}{\cubic\cm}.

The constructed \ac{MMR} consists of two permanent magnets, separated by a distance of approximately \SI{2}{\mm}, called rotator and stator. The rotator is suspended by a thin Dyneema filament (PE-UHMW) that adheres to the spherical magnet (N40, $\SI{4}{\mm}$ diameter) with epoxy resin. The cylindrical stator (N35, $\SI{4}{\mm}$ height and diameter) is fixed to the housing. The attractive magnetic force of the anti-parallel alignment exceeds gravity by several orders of magnitude~\cite{gleich_miniature_2023} and ensures that the rotator does not come into contact with the 3D printed housing (Clear Resin V4, Formlabs, USA). This allows for a low-friction torsional oscillation of the rotator at around \SI{200}{\Hz} for this specific MMR.

The combined DAC/ADC system \ac{RP} is employed for signal generation and data acquisition with a resolution of 14 bits at a maximum rate of 125\,MHz, with two channels per unit operating with a custom FPGA image~\cite{hackelberg_flexible_2022}.
Rx signals are amplified using a single-sided \ac{LNA}, based on the operational amplifier ADA4898-2 (Analog Devices Inc., USA).
A measurement sequence consists of multiple identical frames, where one frame cycle is divided in two distinct windows: the transmit window and the receive window. The values assigned to these windows are $T_\text{tx}=\SI{100}{\ms}$ and $T_\text{rx}=\SI{1900}{\ms}$ respectively, and these are identical for all frames of a sequence (frame rate of \SI{0.5}{\Hz}). Each window has an idle time at the beginning, called the switch time until signals are toggled.  
The measurements are controlled and recorded on a personal computer running a custom measurement software developed in the open-source programming language Julia, connected via Ethernet to the \acp{RP}, shown in \autoref{fig:overview}\,(b).

\subsection{Experiments}
\label{sec:methods:exp}

A total of four measurements are performed to demonstrate the viability of the proposed designs: Low and high \ac{MMR} excitation, each for both \ac{TxRx} switch types. The signals are post-processed to filter out high frequency noise from the environment.
In case of the \ac{DPDT} relay, the amplitude of the class-D amplifier is set to two different levels. A sinusoidal excitation at \SI{200}{\Hz} with \SI{180}{\micro\tesla} and \SI{50}{\micro\tesla} are chosen to provide a visual difference in excitation strength and resulting \ac{MMR} resonance.
For the H-bridge type, the frequency remains the same, but the duty cycle is changed to achieve likewise two different levels of excitation and resonance. Values of $0.8$ and $0.2$ are selected at a constant bridge voltage that results in a peak \SI{180}{\micro\tesla} field. 
Identical coil and \ac{MMR} orientations and positions are used (fixed).

\begin{figure*}[ht!]
    \centering
    \includegraphics[width=1.0\linewidth]{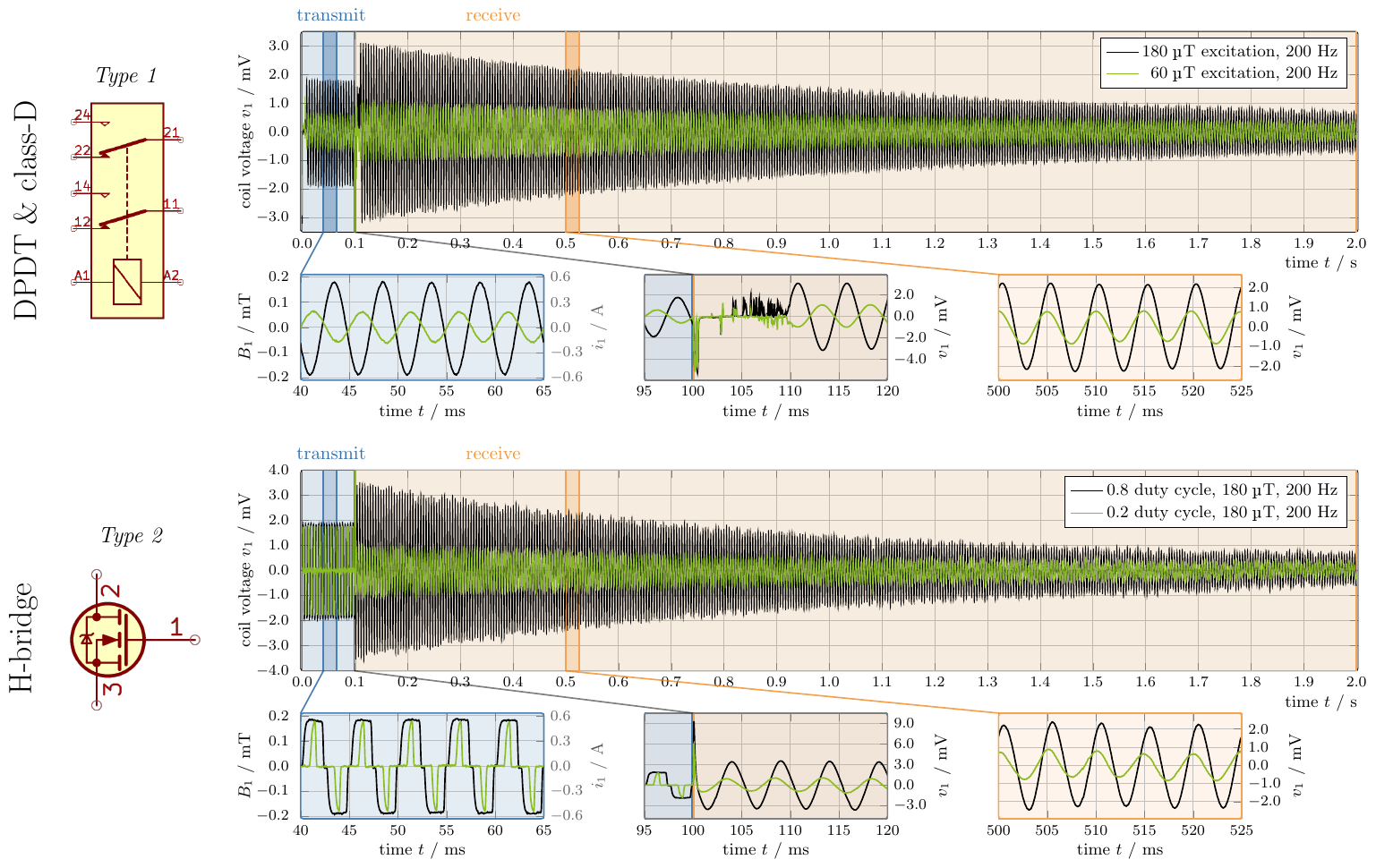}
    \caption{\textbf{MMR measurement results.} A single frame for both proposed TxRx switch types is shown, with superimposed low (green) and high (black) amplitude measurement.
    The sequence cycle of fixed length is divided into $T_\text{tx}=100$\,ms (blue) and $T_\text{rx}=1900$\,ms (orange). The excitation frequency is $200$\,Hz for this specific MMR, consisting out of a spherical and a cylindrical permanent magnet. Frequency, amplitude and phase can be controlled channelwise in real-time between different frames. 
    During excitation, the coil current $i_1$ in $L_1$ is measured and connected to the idle ADC input to provide accurate feedback. Zooms show $5$ Tx and Rx periods of each measurement and the switch window at the beginning of the Rx window. \vspace{-0.2cm}
    }
    \label{fig:results}
\end{figure*}

\section{Results}
\label{sec:results}

\ac{MMR} measurement results are shown in \autoref{fig:results}. For each switch type, two frames are overlaid to visualize low (green) and high (black) amplitude signals. Different zooms show $5$ periods of Tx and Rx signals at \SI{40}{\ms} and \SI{500}{\ms}, and the switch time at \SI{100}{\ms}. The magnetic field $B_1$ value is based on a Biot-Savart simulation\footnote{Package in the Julia programming language. \url{https://github.com/IBIResearch/MagneticFieldConfigurator.jl}} 
to determine the coil sensitivity of the coil pair. 

Both proposed circuits \textit{Type 1} and \textit{Type 2} successfully switch between Tx and Rx windows, generating the desired fields and capturing the decaying resonator signal. Current feedback is incorporated during the Tx window via a sensor on the \ac{PCB}.
Highlighted in the blue zoom areas are the proposed differences in excitation, by adjusting the duty cycle of the pulsed waveform to control the oscillator amplitude. A pulsed current of $i_1=\SI{540}{\mA}$ for $20$ periods with a duty cycle of $0.8$ causes a maximum resonator receive voltage above \SI{3}{\mV}, whereas it is slightly lower for a sinusoidal current at the same amplitude. The selection of a $0.2$ duty cycle results in a decrease of the measured maximum \ac{MMR} resonance by a factor of approximately $3$, similar to the reduction from $180$ to \SI{60}{\micro\tesla} amplitude as seen for the class-D amplifier. We note that the class-D Tx signal for \SI{180}{\micro\tesla} has an intentional \SI{200}{\degree} phase shift.

Although limited to a pulsed waveform, the bridge can drive the \ac{MMR} or LC-resonator to much higher frequencies and amplitudes than low-cost class-D audio amplifiers are able to. Sequence cycle times are reduced and do not depend on relay operating and release times. The results show that the \ac{MMR} signal is visible after \SI{250}{\micro\s} for the H-bridge, while the relay operates at approximately \SI{10}{\ms}. Moreover, the class-D amplifier also requires a delay of about one period for internal control.

\section{Discussion and Conclusion}
\label{sec:discussion}

This work demonstrates the feasibility to build a low-cost analog signal chain for passive resonators with a high quality factor, using either off-the-shelf class-D amplifiers in combination with a \ac{DPDT} relay or a custom H-bridge circuit for pulsed operation. 
Our focus is on \acp{MMR} for prototyping, sensing and localization at oscillator frequencies below \SI{10}{\kHz}~\cite{knopp_empirical_2024,fischer_magneto-oscillatory_2024}.
For higher frequencies and amplitudes, such as those exhibited by the LCQ resonators~\cite{gleich_miniature_2023} at \SI{32}{\kHz}, we propose an H-bridge that integrates \ac{PA} and very fast \ac{TxRx} switching. Both configurations are particularly advantageous for applications requiring a greater number of channels to achieve scalability to a larger field of view. It is crucial to adjust the sequence times to align with the demands of the application and ensure sufficient excitation, as well as an adequate number of receive periods to calculate control parameters for the subsequent frame. Typically, frame rates are below \SI{100}{\ms} in miniature tracking and sensing applications~\cite{gleich_miniature_2023}.

Reasons for using a mechanical \ac{EMR} include the simple circuit design, especially during prototyping stages. The \ac{DPDT} relay switch provides full flexibility in excitation waveform, depending on amplifier specifications and coil impedance. Other amplifier classes, such as A/B types, can be employed. However, current class-D amplifiers are superior in size, cost and efficiency. Our circuit is intended with a non-differential \ac{LNA}. Other circuits variations are possible for differential \acp{LNA} that allow further streamlining of the TxRx circuit.
The main limitation of the \ac{DPDT} relay approach is the maximum sequence frame rate due to the operate and release times of the mechanical relay switch operation. Typically, such \acp{EMR} have an operate time between \SIrange{7}{20}{\ms}~\cite{finder_relays_inc_miniature_2024} and are limited to around \SI{1e6}{cycles}. Their lifetime is further reduced by hot switching, which can be avoided by synchronizing DAC signal and switching to zero-crossings. For higher excitation frequencies the coil inductance may be the limiting factor.

MOSFET switching is in the nanosecond range, allowing very high frame rates and precise control of window size and duty cycle. Although the overall energy efficiency of the circuit is higher than that of \acp{EMR}, the drain-source on resistance is typically higher, adding a series impedance to the coil impedance. 
Due to missing \ac{PWM} control, the waveform is restricted to pulsing in this implementation. In applications that require small bandwidth, for example if multiple \acp{MMR} or LC resonators are tracked simultaneously, this may be a limiting factor where sinusoidal excitation provides a narrow bandwidth. This issue is similar for \acp{MMR} that tend to couple to undesired oscillation modes if excited at specific angles or other/multiple frequencies.

Current feedback can be used for limiting excitation field strength, or as one way to control excitation in combination with the known coil sensitivity.
Our setup can dynamically adjust frequency, amplitude and phase at runtime on the FPGA, however, the sizes of Tx and Rx windows are fixed for all (identical) frames during an ongoing measurement. It is therefore not possible to dynamically change the number of excitation periods, which is essential for pulsed excitation to control resonator amplitudes in maintaining constant oscillation for cases where the resonator undergoes dynamic changes in distance or orientation. This problem is overcome and demonstrated for the pulsed H-bridge by leveraging a run-time-variable duty cycle to compel the same control on resonator amplitude excitation. In contrast, the class-D amplifier based circuit is not susceptible to this issue and can adapt amplifier input amplitudes to achieve the same outcome.

The presented methodology allows the implementation of a multi-channel \ac{TxRx} switch circuit and enables the evaluation of the passive resonator performance in both steady state and transient conditions at different excitation waveforms, frequencies and amplitudes.

\section*{Data Availability Statement}

The data that support the findings of this study are available on zenodo.org under DOI \href{https://doi.org/10.5281/zenodo.14764827}{10.5281/zenodo.14764827}.


\bibliographystyle{Bibliography/IEEEtranTIE}
\bibliography{Bibliography/IEEEabrv,Bibliography/MMRLiterature_zotero}\ 

\begin{thebibliography}{10}
\providecommand{\url}[1]{#1}
\csname url@samestyle\endcsname
\providecommand{\newblock}{\relax}
\providecommand{\bibinfo}[2]{#2}
\providecommand{\BIBentrySTDinterwordspacing}{\spaceskip=0pt\relax}
\providecommand{\BIBentryALTinterwordstretchfactor}{4}
\providecommand{\BIBentryALTinterwordspacing}{\spaceskip=\fontdimen2\font plus
\BIBentryALTinterwordstretchfactor\fontdimen3\font minus
  \fontdimen4\font\relax}
\providecommand{\BIBforeignlanguage}[2]{{%
\expandafter\ifx\csname l@#1\endcsname\relax
\typeout{** WARNING: IEEEtran.bst: No hyphenation pattern has been}%
\typeout{** loaded for the language `#1'. Using the pattern for}%
\typeout{** the default language instead.}%
\else
\language=\csname l@#1\endcsname
\fi
#2}}
\providecommand{\BIBdecl}{\relax}
\BIBdecl

\bibitem{gleich_miniature_2023}
B.~Gleich, I.~Schmale, T.~Nielsen, and J.~Rahmer, ``Miniature
  magneto-mechanical resonators for wireless tracking and sensing,''
  \emph{Science}, vol. 380,
  \href{http://dx.doi.org/10.1126/science.adf5451}{DOI
  10.1126/science.adf5451}, no. 6648, pp. 966--971, Jun. 2023.

\bibitem{fischer_miniature_2024}
F.~Fischer, M.~Jeong, and T.~Qiu, ``Miniature magneto-oscillatory wireless
  sensor for magnetic field and gradient measurements,'' \emph{Applied Physics
  Letters}, vol. 125, \href{http://dx.doi.org/10.1063/5.0222971}{DOI
  10.1063/5.0222971}, no.~7, p. 074102, Aug. 2024.

\bibitem{fischer_magneto-oscillatory_2024}
F.~Fischer, C.~Gletter, M.~Jeong, and T.~Qiu, ``Magneto-oscillatory
  localization for small-scale robots,'' \emph{npj Robotics}, vol.~2,
  \href{http://dx.doi.org/10.1038/s44182-024-00008-x}{DOI
  10.1038/s44182-024-00008-x}, no.~1, p.~1, Mar. 2024.

\bibitem{collins_miniature_1967}
C.~C. Collins, ``Miniature {{Passive Pressure Transensor}} for {{Implanting}}
  in the {{Eye}},'' \emph{IEEE Transactions on Biomedical Engineering}, vol.
  BME-14, \href{http://dx.doi.org/10.1109/TBME.1967.4502474}{DOI
  10.1109/TBME.1967.4502474}, no.~2, pp. 74--83, Apr. 1967.

\bibitem{thanalakshme_magneto-mechanical_2022}
R.~P. Thanalakshme, A.~Kanj, J.~Kim, E.~{Wilken-Resman}, J.~Jing, I.~H.
  Grinberg, J.~T. Bernhard, S.~Tawfick, and G.~Bahl, ``Magneto-{{Mechanical
  Transmitters}} for {{Ultralow Frequency Near-Field Data Transfer}},''
  \emph{IEEE Transactions on Antennas and Propagation}, vol.~70,
  \href{http://dx.doi.org/10.1109/TAP.2021.3137244}{DOI
  10.1109/TAP.2021.3137244}, no.~5, pp. 3710--3722, May. 2022.

\bibitem{knopp_empirical_2024}
T.~Knopp, F.~Mohn, F.~Foerger, F.~Thieben, N.~Hackelberg, J.~Faltinath,
  A.~Tsanda, M.~Boberg, and M.~M{\"o}ddel, ``Empirical {{Study}} of {{Magnet
  Distance}} on {{Magneto-Mechanical Resonance Frequency}},'' \emph{Current
  Directions in Biomedical Engineering}, vol.~10,
  \href{http://dx.doi.org/10.1515/cdbme-2024-2092}{DOI
  10.1515/cdbme-2024-2092}, no.~4, pp. 377--380, Dec. 2024.

\bibitem{kwek_analysis_2008}
L.~B.~K. Kwek, ``Analysis and design of analogue class {{D}} amplifier output
  stages,'' Ph.D. dissertation, Nanyang Technological University, 2008.

\bibitem{tang_design_2020}
B.~Tang, J.~Deng, C.~Zhou, and H.~Ouyang, ``Design of a {{High Efficiency
  Class-D Audio Power Amplifier}},'' in \emph{2020 {{International Conference}}
  on {{Artificial Intelligence}} and {{Electromechanical Automation}}
  ({{AIEA}})}, \href{http://dx.doi.org/10.1109/AIEA51086.2020.00087}{DOI
  10.1109/AIEA51086.2020.00087}, pp. 385--391.\hskip 1em plus 0.5em minus
  0.4em\relax Tianjin, China: IEEE, Jun. 2020.

\bibitem{niklaus_100_2023}
P.~S. Niklaus, J.~W. Kolar, and D.~Bortis, ``100 {{kHz Large-Signal Bandwidth
  GaN-Based}} 10 {{kVA Class-D Power Amplifier With}} 4.8 {{MHz Switching
  Frequency}},'' \emph{IEEE Transactions on Power Electronics}, vol.~38,
  \href{http://dx.doi.org/10.1109/TPEL.2022.3213930}{DOI
  10.1109/TPEL.2022.3213930}, no.~2, pp. 2307--2326, Feb. 2023.

\bibitem{chang_analysis_2000}
J.~Chang, {Meng-Tong Tan}, {Zhihong Cheng}, and {Yit-Chow Tong}, ``Analysis and
  design of power efficient class {{D}} amplifier output stages,'' \emph{IEEE
  Transactions on Circuits and Systems I: Fundamental Theory and Applications},
  vol.~47, \href{http://dx.doi.org/10.1109/81.852942}{DOI 10.1109/81.852942},
  no.~6, pp. 897--902, Jun. 2000.

\bibitem{jockram_concept_2012}
J.~Jockram, O.~Woywode, B.~Gleich, and K.~Hoffmann, ``Concept for a {{Modular
  Class-D Amplifier}} for {{MPI Drive Field Coils}},'' in \emph{Magnetic
  {{Particle Imaging}}}, vol. 140, pp. 343--347.\hskip 1em plus 0.5em minus
  0.4em\relax Berlin, Heidelberg: Springer Berlin Heidelberg, 2012.

\bibitem{hurtado-velasco_simulation_2016}
R.~{Hurtado-Velasco} and J.~{Gonzalez-Llorente}, ``Simulation of the magnetic
  field generated by square shape {{Helmholtz}} coils,'' \emph{Applied
  Mathematical Modelling}, vol.~40,
  \href{http://dx.doi.org/10.1016/j.apm.2016.06.027}{DOI
  10.1016/j.apm.2016.06.027}, no. 23-24, pp. 9835--9847, Dec. 2016.

\bibitem{hackelberg_flexible_2022}
N.~Hackelberg, J.~Schumacher, M.~Graeser, and T.~Knopp, ``A {{Flexible
  High-Performance Signal Generation}} and {{Digitization Plattform}} based on
  {{Low-Cost Hardware}},'' \emph{International Journal on Magnetic Particle
  Imaging}, \href{http://dx.doi.org/10.18416/IJMPI.2022.2203063}{DOI
  10.18416/IJMPI.2022.2203063}, p. Vol 8 No 1 Suppl 1 (2022), Mar. 2022.

\bibitem{finder_relays_inc_miniature_2024}
{Finder Relays inc.}, ``Miniature {{PCB Relays}}, 40 series,''
  https://cdn.findernet.com/app/uploads/S40EN.pdf, 2024.

\end{thebibliography}

\end{document}